\documentclass[aps,pra,amsmath,superscriptaddress,reprint,floatfix,notitlepage,balancelastpage,twocolumn,reprint]{revtex4}
\usepackage{graphicx}% Include figure files
\usepackage{bm,color,subfigure}
\newcommand{\nf}{n_{\rm F}}

\newcommand{\Tr}{{\rm Tr}}

\newcommand{\kb}{k_{\rm B}}

\newcommand{\ph}{\hat{p}}

\newcommand{\be}{\begin{equation}}
\newcommand{\ee}{\end{equation}}
\newcommand{\bea}{\begin{eqnarray}}
\newcommand{\eea}{\end{eqnarray}}
\newcommand{\bse}{\begin{subequations}}
\newcommand{\ese}{\end{subequations}}

\input{epsf}
\begin{document}
%\draft
\title{Microcanonical analysis of Boltzmann and Gibbs Entropies in trapped cold atomic gases}
%-------------------------
\author{Kenneth J.~Higginbotham}
\affiliation{School of Physics, Georgia Institute of Technology, Atlanta, Georgia 30332}
\author{Daniel E.~Sheehy}
\affiliation{Department of Physics and Astronomy, Louisiana State University, Baton Rouge, Louisiana 70803}
\date{\today}
%\maketitle
%--------------------------
\begin{abstract}
%--------------------------
We analyze a gas of noninteracting fermions confined to a one-dimensional  harmonic oscillator potential, with the aim of
distinguishing between two proposed definitions of the thermodynamic entropy in the microcanonical ensemble, namely
the standard Boltzmann entropy and the Gibbs (or volume) entropy.  The distinction between these two definitions is
crucial for systems with an upper bound on allowed
energy levels, where the Boltzmann definition can lead to the notion of negative absolute temperature.  Although negative temperatures
do not exist for the system of fermions studied here, we still find a significant difference between the Boltzmann and Gibbs entropies,
 and between the corresponding temperatures with the Gibbs temperature being closer (for small particle number) to the
temperature based on a grand canonical picture.   
%
%   We propose that cold atom experiments on small numbers of fermions
% in a 1D trap could discern which definition is closer to the true 
% The distinction between these definitions is
% closely tied to the notion of negative temperature, which
%
%--------------------------
\end{abstract}
\maketitle
%--------------------------
%
\section{Introduction}
Recent work by Dunkel and Hilbert~\cite{DunkelNatPhys} (DH), 
motivated by classic early experiments on spin systems~\cite{Purcell1951,Ramsay1956} and recent experiments on cold atomic 
gases~\cite{Braun2013} confined to optical lattices,
 has proposed that the notion of negative absolute temperatures 
 arises from a definition of entropy that is thermodynamically inconsistent.

The idea of negative temperatures has existed at least since the 1950's, when it was applied to 
understand experiments on spin systems~\cite{Purcell1951,Ramsay1956}, and
since then it has become standard textbook material in thermodynamics and statistical physics.  
 More recently, evidence of negative temperature
has been found in experiments on cold atomic gases~\cite{Braun2013}.   In each
case, negative temperatures arise because the system in question possesses an upper bound in
the total energy $E$.  For energies close to the upper bound, the number of available microstates 
(which is directly related to the Boltzmann entropy)
decreases with increasing energy, implying a negative absolute temperature using the Boltzmann 
entropy definition.

%----------------------------
\begin{figure}[ht!]
     \begin{center}
        \subfigure{
%            \label{fig:first}
            \includegraphics[width=90mm]{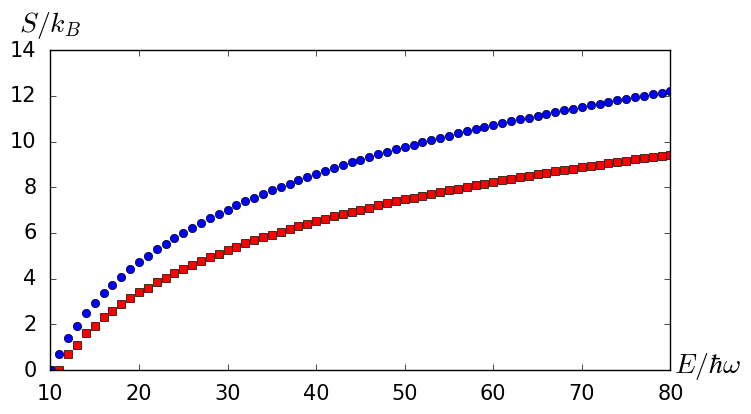}}
        \\ \vspace{-.5cm}
        \subfigure{
 %           \label{fig:fourth}
            \includegraphics[width=90mm]{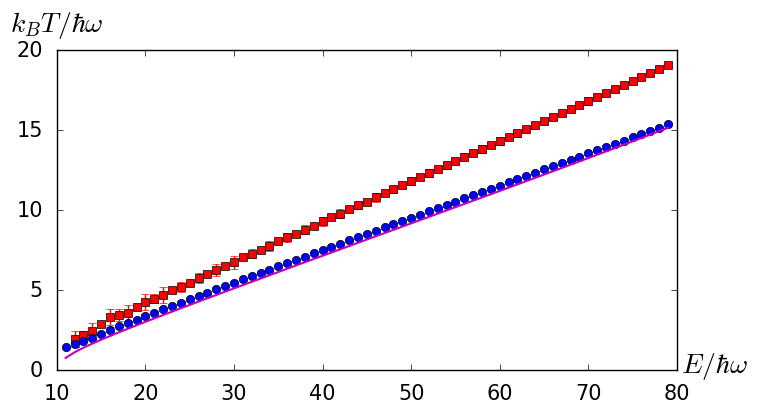}}
    \end{center}\vspace{-.5cm}
    \caption{
        (Color Online) The top panel compares the Gibbs ($S_G$, blue circles) and Boltzmann ($S_B$, red squares) entropy definitions for
a trapped 1D fermionic atomic gas with $N = 5$ fermions as a function of energy $E$ normalized to the oscillator energy.  The bottom panel
compares the Gibbs ($T_G$, blue circles) and Boltzmann temperatures    ($T_B$, red squares), showing a clear difference that increases
with increasing $E$.  The purple line in the bottom panel is the grand-canonical temperature $T_{GC}$ that is obtained assuming
the grand-canonical number and energy equations [Eqs.~(\ref{Eq:numTMu})] hold.  The small error bars on the $T_G$ and
$T_B$ results indicate the uncertainty in evaluating the derivative Eq.~(\ref{eq:tempdef}) given that we only have
entropy values at discrete values of $E$.
     }
\vspace{-.5cm}
   \label{fig:one}
\end{figure}
%----------------------------

For a quantum system defined by Hamiltonian $H$, the Boltzmann entropy can be written as 
\be
\label{boltzmannentropy}
S_B(E) =  \kb \ln \big[\varepsilon \Omega(E)\big],
\ee
where $\kb$ is Boltzmann's constant and $\Omega(E) = \Tr\big[ \delta(E-H)\big]$ counts the
number of microstates at energy $E$.  Here, $\varepsilon$ is a parameter with units of energy
chosen so that the argument of the logarithm is dimensionless.   In Ref.~\cite{DunkelNatPhys},
DH argue for an alternate ``volume'' definition of entropy (due to Gibbs)
that instead counts all microstates up to energy $E$.  We can define the Gibbs entropy as:
\be
\label{eqessg}
S_G(E) = \kb \ln \big[ \Tr \Theta(E-H)\Big],
\ee 
with $\Theta$ the Heaviside step function.   Although less well-known, the Gibbs entropy has
appeared in some thermodynamics textbooks (e.g., Ref.~\cite{Becker}).  For either case, the 
temperature is defined by the usual relation
\be
\label{eq:tempdef}
\frac{1}{T} = \frac{\partial S}{\partial E},
\ee
and we thus define the Boltzmann ($T_B$) and Gibbs ($T_G$) temperatures by combining Eq.~(\ref{boltzmannentropy}) or Eq.~(\ref{eqessg}) with
Eq.~(\ref{eq:tempdef}).   For spin systems in magnetic fields, as well as the system of cold atomic gases in optical lattices
studied in Ref~\cite{Braun2013},
the difference in these entropy definitions leads to a situation in which $T_B$ can be negative, while $T_G$ is
positive.  The proposal by DH that the Gibbs definition is the correct one has led to a spirited debate in the 
literature~\cite{SchneiderComment,Dunkel2014p,FrenkelWarren,Dunkel2014,Hilbert2014,Swendsen2015,Poulter,Campisi,Abraham}.
Here, we mainly bypass this debate, and investigate a system (inspired by experiments on cold atomic gases) that does 
not exhibit negative Boltzmann temperature, but which, nonetheless, will have a small (but possibly observable) difference between the
two entropy and temperature definitions. 

\section{Main Results}

In this paper we study the distinction between Gibbs and Boltzmann entropies and temperatures in the context
of ultracold trapped atomic gases, which are perhaps the simplest system that can truly be taken to be in
the microcanonical ensemble.  We study $N$ identical fermionic atoms confined in a quasi-one dimensional trapping potential
(realized by a trap that exhibits tight confinement in two directions and weak confinement in the third).  
Of course, this system does not have an upper energy bound, and therefore is not expected to exhibit negative
temperature.  Nonetheless, there is still a difference between the Gibbs and Boltzmann entropy definitions,
implying a difference between $T_B$ and $T_G$ that could be experimentally significant at small $N$.  Since 
experiments on cold atomic gases have already accessed the small-$N$  regime~\cite{Serwane}
and have realized the quasi-1D regime using an optical lattice potential~\cite{LiaoNature2010,Revelle2016}, we propose that
our results may be relevant for 
future experiments that can distinguish between the Boltzmann and Gibbs temperature definitions, shedding
further light on this controversy.

%----------------------------
%----------------------------
\begin{figure}[ht!]
     \begin{center}
        \subfigure{
%            \label{fig:first}
            \includegraphics[width=90mm]{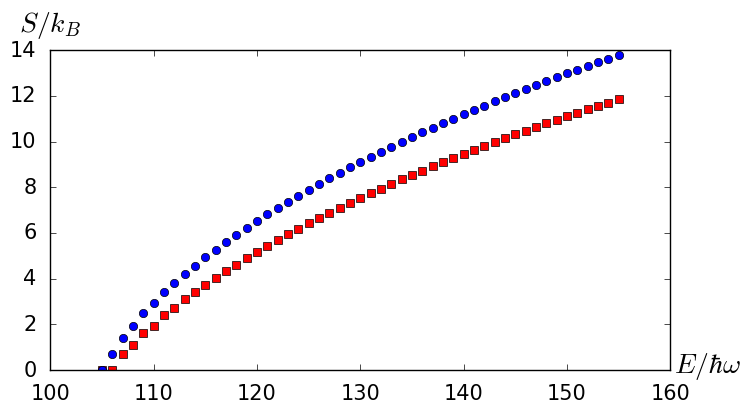}}
        \\ \vspace{-.5cm}
        \subfigure{
 %           \label{fig:fourth}
            \includegraphics[width=90mm]{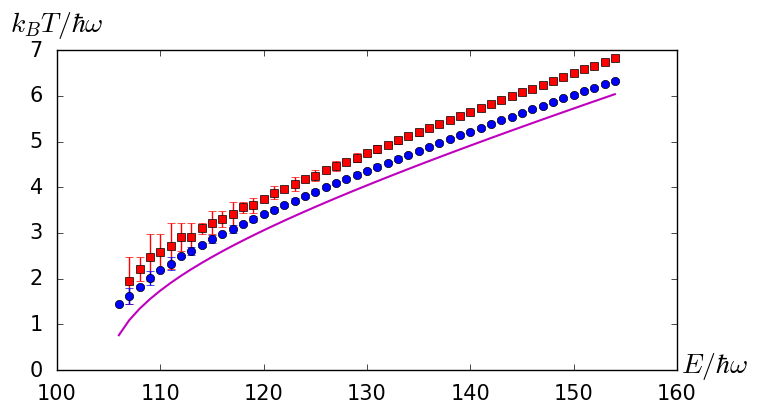}}
    \end{center}\vspace{-.5cm}
    \caption{
         (Color Online)  As in Fig.~\ref{fig:one}, the top panel compares the Gibbs ($S_G$, blue circles) and Boltzmann ($S_B$, red squares) entropy definitions for
a 1D fermionic atomic gas with $N = 15$ fermions as a function of energy $E$ normalized to the oscillator energy $\hbar \omega$.  The bottom panel
compares the Gibbs ($S_G$, blue circles) and Boltzmann temperatures    ($T_B$, red squares), showing a modest difference that increases
with increasing $E$ (as in the $N=5$ case, Fig.~\ref{fig:one}).  The purple line in the bottom panel is the grand-canonical temperature $T_{GC}$ that is obtained assuming
the grand-canonical number and energy equations [Eqs.~(\ref{Eq:numTMu})] hold. 
     }
\vspace{-.5cm}
   \label{fig:two}
\end{figure}
%----------------------------

In asking the question of which temperature definition, Boltzmann or Gibbs, is correct, 
we need a third temperature scale to compare to.  A common way to measure the temperature in cold
atomic gas experiments is by measuring the atom density vs. position and comparing to a theoretical 
formula based on the Fermi distribution~\cite{Hung2011}.  Inspired by this, we shall define a third temperature scale, 
$T_{GC}$,  based on the grand-canonical picture in which single-particle levels are occupied according to the 
Fermi distribution. 
%
 %the  temperature that is extracted in this manner
 %the grand-canonical temperature $T_{GC}$ since it assumes that the grand-canonical picture holds. 
%
 One  question that we shall address is whether the expected value of $T_{GC}$ is closer to $T_B$ or $T_G$
(if either)  for a 1D trapped fermionic atomic gas assumed to be in the microcanonical 
ensemble at total particle number $N$ and energy $E$.   A natural objection to this
line of reasoning is that $T_{GC}$ only holds rigorously for large $N$,
in the thermodynamic limit (where number and energy fluctuations can be neglected), whereas we consider 
the regime of small $N$ (where the difference between $T_B$ and $T_G$ is largest).
 However, in practice we find that the microcanonical density profile is very accurately fit by the grand canonical picture 
even in the small $N$ regime.  This implies that an experiment attempting to extract the temperature by observing
the local density as a function of position would ``measure'' a temperature close to $T_{GC}$.

Our main result can be seen in Fig.~\ref{fig:one} that compares these entropy and temperature 
definitions for the case of $N=5$ fermions.   This figure shows that, at least in the small $N$ regime, the inequalities $S_G>S_B$ 
and $T_{GC}\simeq T_G<T_B$ hold. Indeed, the difference between $T_G$ and $T_B$  increases with increasing energy, while $T_{GC}$ remains
close to $T_G$ suggesting that,
at least at small $N$, the Gibbs definition is the appropriate one (i.e., closer to the definition consistent
with a grand-canonical picture based on the Fermi distribution).  
As discussed below, the difference between $T_G$ and $T_B$ decreases with increasing $N$ (as seen in Fig.~\ref{fig:two}), consistent with the 
expectation that they should be equal in the  thermodynamic limit.  We argue below, however, that for any fixed $N$, for sufficiently large energy, the
qualitative behavior shown in Fig.~\ref{fig:one} holds.

\section{System Hamiltonian and Entropy Calculations}

We study a single-species gas of 
atomic fermions confined to a harmonic trapping potential that is anisotropic, satisfying  $\omega_y = \omega_z \gg\omega_x$.
At sufficiently low numbers of particles, such that the system chemical potential is also much less than $\hbar\omega_z$, such a gas can be
accurately modeled by the one-dimensional second quantized Hamiltonian
\be
\label{Eq:ham}
H =  \int dx \, \Psi^\dagger(x) \Big[ \frac{\ph^2}{2m} + \frac{1}{2}m\omega^2 x^2\Big]\Psi(x) ,
\ee
where $\ph = -i\hbar \frac{d}{dx}$ is the momentum operator, $m$ the particle mass, and  the fermionic field operators  
$\Psi(x)$ satisfy the anticommutation relation 
$\{ \Psi(x) , \Psi^\dagger(x') \} = \delta(x-x')$.  It is convenient to express $H$ in terms of
mode operators $c_n$, which are related to the field operators by
\be
\label{Eq:spivc}
\Psi(x) = \sum_{n=0}^\infty \psi_n(x) c_n, 
\ee
where $\psi_n(x)$ is the well-known solution to the one-dimensional quantum harmonic oscillator, 
\be
\psi_n(x) = 
\frac{1}{\sqrt{2^n n!}a^{1/4}} \frac{1}{\pi^{1/4}} {\rm e}^{-\frac{x^2}{2a^2}} H_n(x/a),
\ee
where $H_n(x)$ is the $n$th Hermite polynomial and $a = \sqrt{\frac{\hbar}{m\omega}}$ is the oscillator length.  The
corresponding eigenvalue is $\epsilon_n = \hbar \omega(n+\frac{1}{2})$, giving for the system Hamiltonian, after 
plugging Eq.~(\ref{Eq:spivc}) into Eq.~(\ref{Eq:ham}), 
\be
H = \sum_{n=0}^\infty \epsilon_n c_n^\dagger c_n.
\ee
Henceforth, we shall measure lengths in units of the oscillator length and measure energies relative to $\hbar \omega$. 
 It is also
convenient to drop the zero point energy, so that $\epsilon_n = n$.
Our next task is to analyze the behavior of a gas of $N$ fermions, described by $H$, in the microcanonical ensemble.   Then,
a member of this ensemble is described by a wavefunction with certain levels occupied:
\be
|\Psi\rangle = c_{n_1}^\dagger c_{n_2}^\dagger \cdots c_{n_N}^\dagger |0\rangle, 
\ee
with $|0\rangle$ the vacuum state. Due to the Pauli principle, no two $n_i$ may
coincide.  

Since the energy eigenvalues are discrete, our definition of $\Omega(E,N)$ will differ
from that given below Eq.~(\ref{boltzmannentropy}) but instead simply count the total 
number of allowed
microstates with total energy $E$.  Indeed,  
since the single-particle energies are integers, the energy eigenvalue of $|\Psi\rangle$,
$E[\{n_i\}] = \sum_{i=1}^N n_i$, is also an integer, so that the problem of determining 
$\Omega(E,N)$  is related to the well-known integer partitioning problem.  Following
standard notation, we define $P(n)$ to be the total number of partitions of the integer $n$ and $P(n,m)$
to be the number of partitions of  the integer $n$ into exactly $m$ parts so that, for example,
$P(3) = 3$ (with the partitions being $\{\{3\}, \{2,1\}, \{1,1,1\}\}$)  and $P(5,3)= 2$ 
(with the partitions being $\{\{3,1,1\}, \{2,2,1\}\}$).  Both functions allow repetitions of 
integers appearing in the
partitions, which we must exclude due to the Pauli principle.  However, it turns out that 
the number of partitions of the integer $n$ into  exactly $m$ parts excluding 
repetitions, $Q(n,m)$, is related to $P(n,m)$ by:
\be
\label{eq:qvp} 
Q(n,m) = P(n- \frac{1}{2}m(m-1),m).
\ee
To verify this well-known identity, we establish a one-to-one correspondence between partitions associated with
the left and right sides of this formula.  Thus,
consider one of the partitions on the right side, which is of the form
$\{n_1,n_2\cdots ,n_{m-1}, n_m\}$.  By definition, this partition has energy $E[\{n_i\}] = n -  \frac{1}{2}m(m-1)$ and may
contain some number of repeated integers but satisfies $n_i\geq n_{i+1}$ (with larger integers to the left as in
the above examples).  However, a new partition with no repeated
integers can be obtained by incrementing the integers in this partition thusly: $\{n_1+m-1, n_2+m-2,\cdots, n_{m-1}+1,n_m\}$.
This partition clearly satisfies $n_i>n_{i+1}$, and has energy $n$ since the increments of each integer add to $\frac{1}{2}m(m-1)$.
Since it is clear that any restricted partition (corresponding to $Q(n,m)$) can be connected to an unrestricted partition (corresponding to
 $P(n- \frac{1}{2}m(m-1),m)$), the relation Eq~(\ref{eq:qvp}) holds.  To obtain our final expression for  $\Omega(E,N)$ in terms of $Q(n,m)$,
we recall that a fermion in the lowest level $n=0$ has zero energy (not contributing to $E$).
This finally implies that the total number of microstates has two contributions:
\be
\label{Eq:microstates}
\Omega(E,N) = Q(E,N) + Q(E,N-1), 
\ee 
with the first (second) term on the right side corresponding to microstates in which the 
$n=0$ level is unoccupied (occupied).

%----------------------------
\begin{figure}[ht!]
     \begin{center}
 \includegraphics[width=85mm]{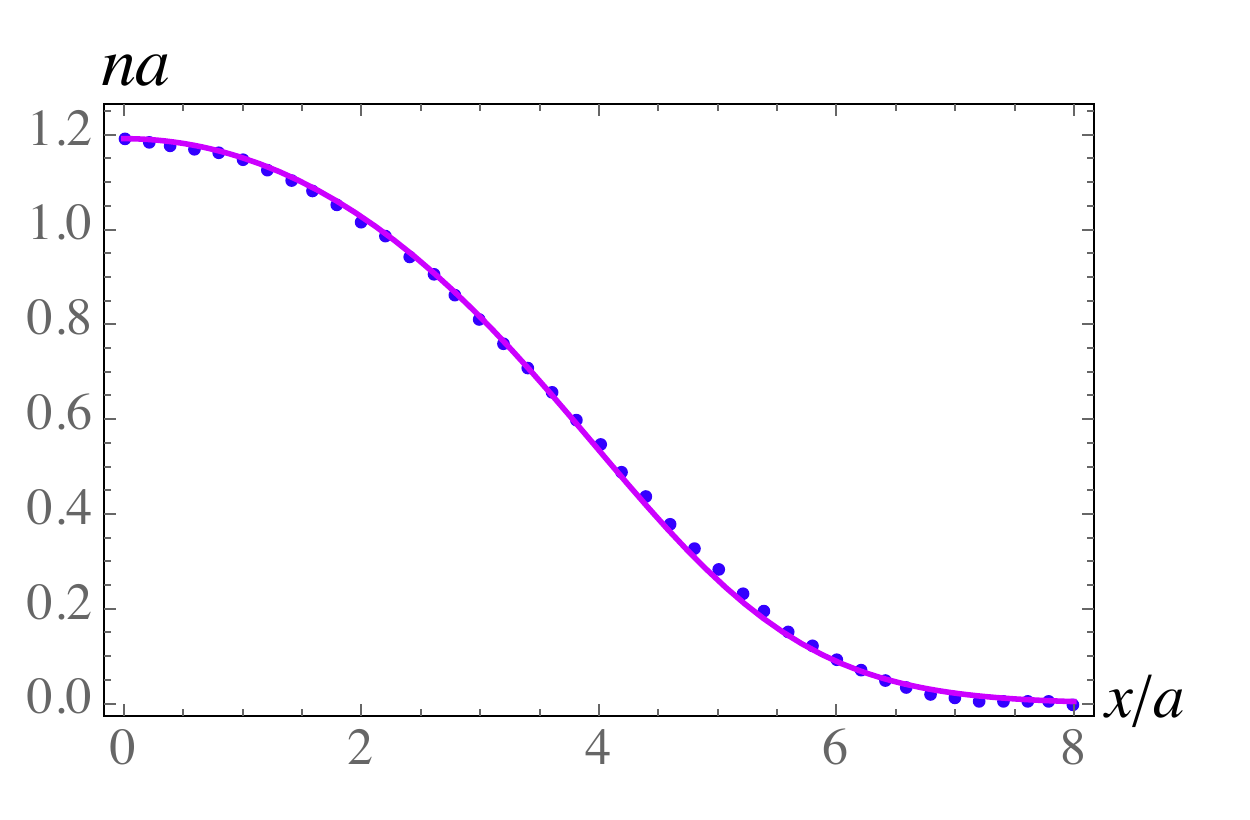}
%
%        \subfigure{
%            \label{fig:first}
%            \includegraphics[width=85mm]{Fig2.pdf}}
%        \\ \vspace{-.5cm}
%        \subfigure{
%            \label{fig:fourth}
%            \includegraphics[width=85mm]{Fig1b.pdf}}
    \end{center}\vspace{-.5cm}
    \caption{(Color Online)
      Plot of local density $n$ (normalized to $a^{-1}$ with $a$ the oscillator length) vs. position $x$ (normalized to $a$) 
for a system with $N=9$ particles
and energy $E = 60 \hbar\omega$, comparing the microcanonical density according to Eq.~(\ref{Eq:mcd})
(solid points) to the grand-canonical density (solid purple line) Eq.~(\ref{Eq:gcd}).
     }
\vspace{-.25cm}
   \label{fig:densityfigure}
\end{figure}
%----------------------------

In terms of Eq.~(\ref{Eq:microstates}), the Boltzmann entropy is 
\be
S_B(E)  = \kb \ln \Omega(E,N), 
\ee
while the Gibbs entropy sums over all allowed total energies less than $E$:
\be
\label{eq:ged}
S_G(E) = \kb \ln \sum_{E'<E} \Omega(E',N).
\ee
In Figs.~\ref{fig:one} and \ref{fig:two}, we show our numerical results for $S_B$ and $S_G$  (top
panel) along with the Boltzmann and Gibbs temperatures (bottom panel) as a function of the
total system energy $E$.   Due to the Pauli principle, the minimum system energy
is 
\be
\label{eq:emin}
E_{\rm min} = \frac{1}{2}N(N-1),
\ee
 We see that 
the Gibbs entropy is larger than the Boltzmann entropy, $S_G>S_B$, with a difference
that increases with increasing system energy.  Similarly, the Gibbs
temperature (obtained by numerically differentiating the entropy results and using Eq.~(\ref{eq:tempdef}))
 is lower than the Boltzmann temperature.  
The solid line in the bottom panel of   Figs.~\ref{fig:one} and \ref{fig:two} shows
 the ``grand-canonical''  temperature $T_{GC}$, extracted by  assuming that the equations for
the grand-canonical
ensemble hold:
\bse
\label{Eq:numTMu}
\bea
\label{Eq:numMu}
N &=& \sum_{n=0}^\infty \nf(\epsilon_n -\mu),
\\
E &=& \sum_{n=0}^\infty \epsilon_n \nf(\epsilon_n -\mu) ,
\eea
\ese
with $\nf(x) = \frac{1}{{\rm e}^{\beta x}+1}$ (and $\beta = \frac{1}{\kb T}$) the Fermi distribution.
Although these only determine the mean particle number and energy (with fluctuations that vanish 
in the thermodynamic limit), they provide a unique prediction that can be compared to
$T_B$ and $T_G$.  

We now argue that the local density profile of a 1D trapped fermionic gas in the microcanonical ensemble is
approximately consistent with the grand canonical ensemble picture even at small $N$.  To establish this, 
 in Fig.~\ref{fig:densityfigure}, we compare the microcanonical density, 
\be
\label{Eq:mcd}
n(x) = \frac{1}{\Omega(E,N)} \sum_{\{n_i\}} \sum_{i=1}^N |\psi_{n_i}(x)|^2,
\ee
for the case of $E=60$ and $N= 9$, to the grand canonical density 
\be
\label{Eq:gcd}
n_{GC}(x) = \sum_{n=0}^\infty \nf(E_n -\mu)|\psi_n(x)|^2,
\ee
using the temperature ($T_{GC} = 4.4$) and chemical potential ($\mu_{GC} = 7.8$) 
obtained by solving Eqs.~(\ref{Eq:numTMu}) for the same system parameters.  
For these parameters, $T_G\simeq 4.8$ while $T_B \simeq 5.5$.
 In cold atomic
gas experiments, the temperature is often extracted by measuring the density profile (and assuming the 
grand-canonical picture holds).  Thus, the close agreement in Fig.~\ref{fig:densityfigure}
suggests that a cloud at this energy and particle number would be ``measured'' to have a 
temperature given by $T_{GC}$, closer to $T_G$.

%----------------------------
\begin{figure}[ht!]
     \begin{center}
 \includegraphics[width=85mm]{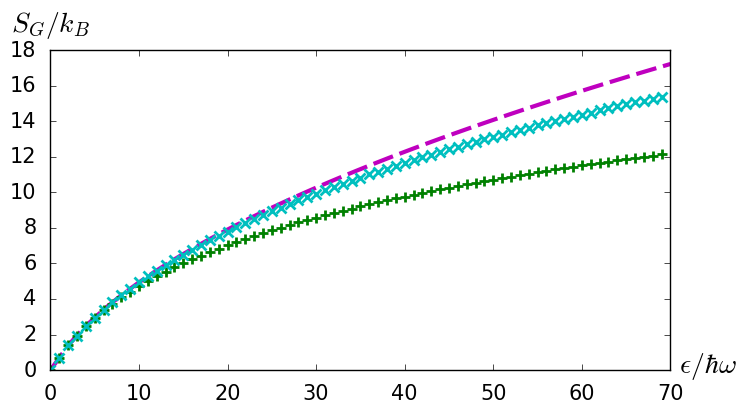}
%
%        \subfigure{
%            \label{fig:first}
%            \includegraphics[width=85mm]{Fig2.pdf}}
%        \\ \vspace{-.5cm}
%        \subfigure{
%            \label{fig:fourth}
%            \includegraphics[width=85mm]{Fig1b.pdf}}
    \end{center}\vspace{-.5cm}
    \caption{(Color Online)
     Gibbs entropy, normalized to $\kb$, as a function of the energy relative to the minimum energy ($\epsilon = E- E_{\rm min}(N)$), 
for the cases of $N=5$ (green plusses) and $N=10$ (blue crosses).  They agree with each other and with the universal 
entropy (following from plugging Eq.~(\ref{eq:omegalowepsilon}) into Eq.~(\ref{eq:ged})) (dashed purple) at low $\epsilon$ and diverge at large $\epsilon$.
 }
\vspace{-.25cm}
   \label{fig:universalfigure}
\end{figure}
%----------------------------

\section{Universal entropy at low energies}

Our results suggest that the Gibbs and Boltzmann
entropy and temperature definitions agree at small $E$ but differ at large $E$, with the Gibbs
temperature definition being closer to the grand-canonical temperature.  In this section we show that
the Gibbs and Boltzmann entropy formulas have a universal form at low energies that greatly 
simplifies their calculation.  We will propose that the approximate agreement between $T_B$ and $T_G$
is related to the fact that the universal formulas hold at low energies, and that the deviation of 
$T_B$ from $T_G$ occurs when the system is outside of the universal regime.

%
%shows that, while the three temperatures agree
%for   for small system energies  (i.e., $T_B\simeq T_G\agt T_{GC}$),
%at larger system energies the Boltzmann temperature begins to deviate
%from the Gibbs temperature (but $T_G\agt T_{GC}$ still holds).  We find that, with increasing $N$, the
%energy range over which the Boltzmann and Gibbs approximately coincide
%increases.  

To establish the existence of the universal entropy formulas, we define the relative 
$\epsilon = E - E_{\rm min}$ with $E_{\rm min}$ the minimum energy
Eq.~(\ref{eq:emin}).  Then, as illustrated in Fig.~\ref{fig:universalfigure}, the Gibbs entropies, 
plotted as a function of $\epsilon$, are universal  (independent of $N$) for sufficiently small $\epsilon$.
A similar universality holds for $S_B$.

%----------------------------
\begin{figure}[ht!]
     \begin{center}
        \subfigure{
%            \label{fig:first}
            \includegraphics[width=85mm]{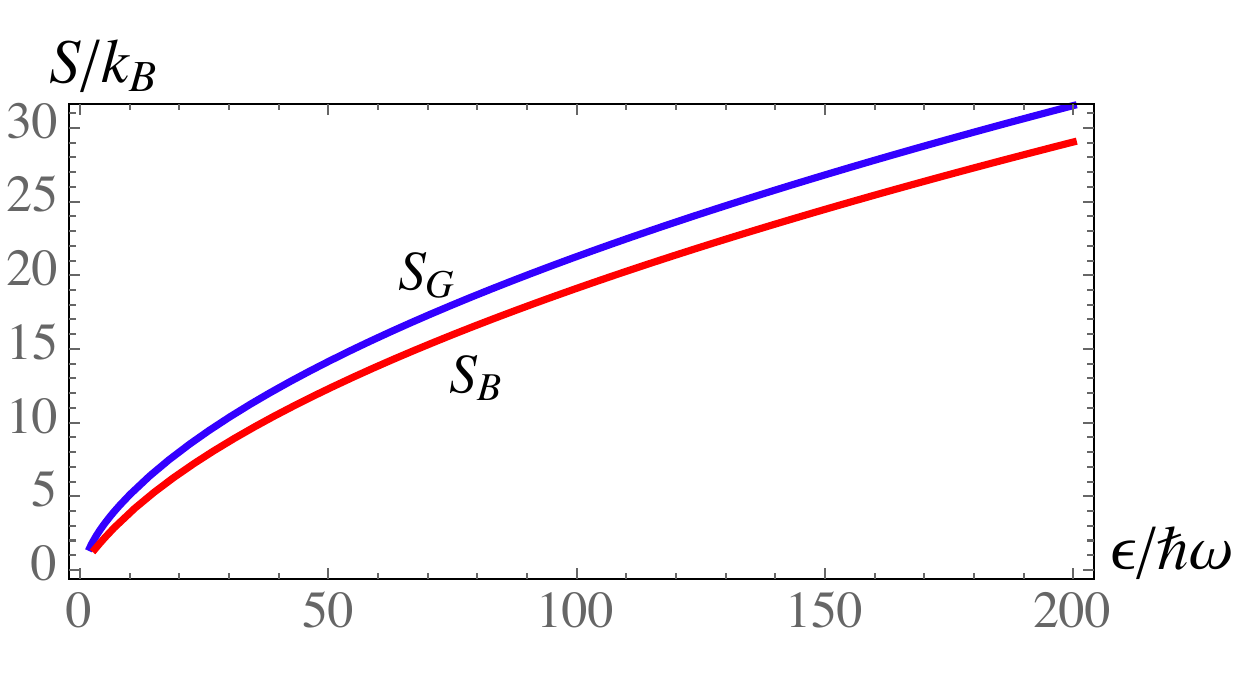}}
        \\ \vspace{-.5cm}
        \subfigure{
 %           \label{fig:fourth}
            \includegraphics[width=85mm]{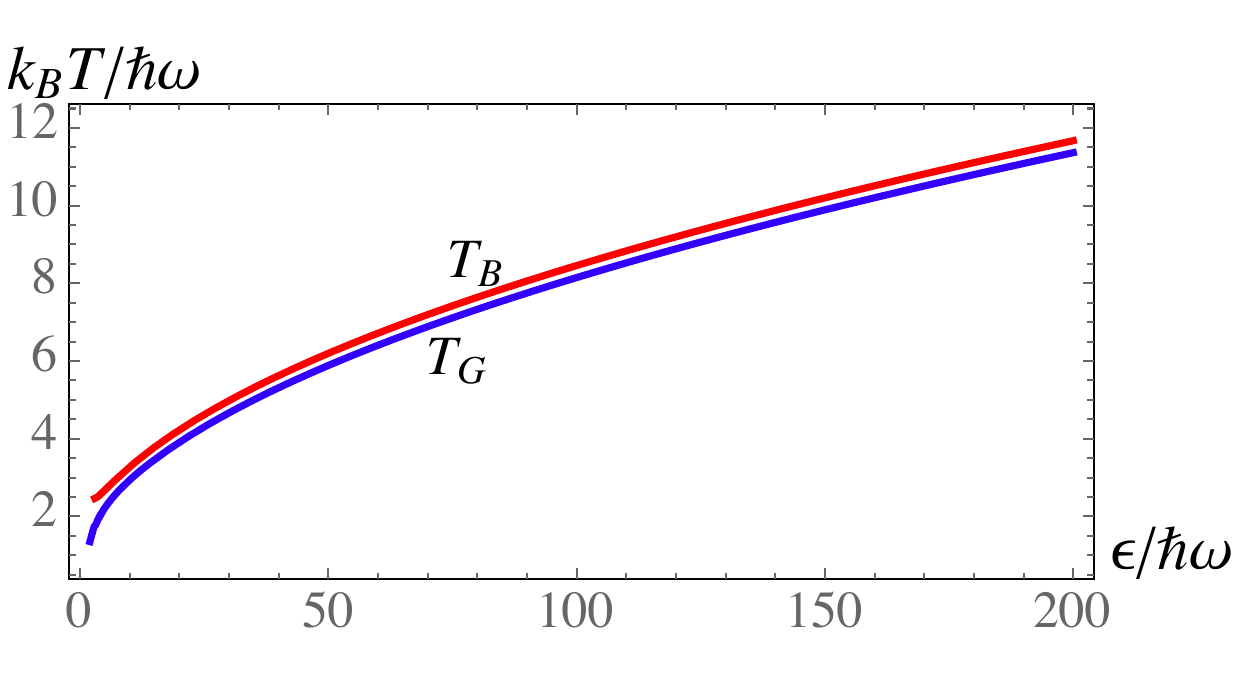}}
    \end{center}\vspace{-.5cm}
    \caption{
        (Color Online)  The top panel shows the Gibbs and Boltzmann entropies, normalized to the Boltzmann constant,
as a function of system energy using the  formulas Eqs.~(\ref{Eq:bgentropies}) (valid for sufficiently low
$\epsilon  = E- E_{\rm min}$) along with the Hardy-Ramanujan asymptotic approximation for the integer partition function (Eq.~(\ref{eq:hr})). 
 The bottom panel shows the corresponding Gibbs and Boltzmann temperatures, obtained
by numerical differentiation of the results in the top panel along with Eq.~(\ref{eq:tempdef}).
     }
\vspace{-.5cm}
   \label{fig:HardyFormulaFig}
\end{figure}
%----------------------------

This universality at small $\epsilon$ follows from the following mathematical
identity for the integer partition function~\cite{Hwang}:
\be
\label{Eq:hwangresult}
P(n,m) = P(n-m) \hspace{1cm} {\rm for}\,\, m\geq \frac{n}{2},
\ee
relating $P(n,m)$ to the unrestricted partition function at small $n$. When this result
is combined with Eq.~(\ref{eq:qvp}) and plugged into Eq.~(\ref{Eq:microstates}), we find
\be
\label{eq:omegalowepsilon}
\Omega(E,N)  = P(\epsilon) \hspace{1cm} {\rm for}\,\, \epsilon \leq N-1,
\ee
where we note that only the second term of Eq.~(\ref{Eq:microstates}) contributes in 
this small-$\epsilon$ regime.  In this context, the  identity Eq.~(\ref{eq:omegalowepsilon})
can be related, physically, to 
the
Pauli principle: for $\epsilon = 0$, the only allowed microstate of the $N$ 
fermion gas is the ground state which has all levels filled up to the Fermi level $\nf = N-1$.   If we increase
the energy of our system to small $\epsilon$, the allowed microstates simply correspond
to particle-hole excitations of this ground state in which fermions are promoted from levels slightly
below $\nf$ to slightly above $\nf$.  In these microstates, all levels up to level
$m=N-\epsilon-1$ (at least) are occupied.  This follows because exciting a fermion from a lower
level would cost too much energy, and implies that the number of microstates is given by
the number of ways to construct such excitations i.e., by the number of {\em unrestricted} partitions of $\epsilon$.
This implies Eq.~(\ref{eq:omegalowepsilon}).  This picture holds for the microstates only for $m\geq 0$,
implying the restriction $\epsilon \leq N-1$.

 Equation~(\ref{eq:omegalowepsilon})
implies the following expressions for the Boltzmann and Gibbs entropies in the universal regime:
\bse
\bea
S_B(\epsilon) &=& \kb \ln P(\epsilon) , \hspace{0.25cm}  {\rm for}\,\, 0<\epsilon\leq N-1,
\\
S_G(\epsilon) &=& \kb \ln \sum_{\epsilon'=0}^\epsilon P(\epsilon'), \hspace{0.25cm}  {\rm for}\,\, 0<\epsilon\leq N-1.
\eea
\label{Eq:bgentropies}
\ese
The calculation of $S_B$ and $S_G$ via Eq.~(\ref{Eq:bgentropies}) is much easier than via the 
direct formula Eq.~(\ref{Eq:microstates}) since they depend only on  the unrestricted integer partition function.
In fact, a  convenient asymptotic large-$\epsilon$ formula for $P(\epsilon)$ has been derived by 
Hardy and Ramanujan~\cite{Hardy}
\be
\label{eq:hr}
P(\epsilon) \simeq \frac{1}{4\epsilon \sqrt{3}}{\rm e}^{\pi\sqrt{\frac{2\epsilon}{3}}},
\ee
which allows a straightforward approximate numerical evaluation of $S_G$ and $S_B$ in 
the large $\epsilon$ regime.  In the top panel
Fig.~\ref{fig:HardyFormulaFig}, we compare $S_G$ and $S_B$ computed  using Eq.~(\ref{Eq:bgentropies}) 
along with Eq.~(\ref{eq:hr}), with the comparison between $T_G$ and $T_B$ appearing in the bottom panel.  
The close agreement between these curves is natural, given that we expect the difference between the 
Gibbs and Boltzmann entropies to vanish in the large system ``thermodynamic'' limit.    
We propose that, since the univesal formulas Eq.~(\ref{Eq:bgentropies}) only hold for $\epsilon<N-1$, 
the large difference between $S_B$ and $S_G$ that we find for small $N$ occurs when these systems 
are outside of the universal regime and that, for any fixed $N$, the approximate agreement between
$S_B$ and $S_G$ holds only for $\epsilon<N-1$, with differences between $S_B$ and $S_G$ (and between
$T_G$ and $T_B$) occuring for large $\epsilon$.

% agreement only holds in the universal regime $\epsilon \leq N-1$.  
%Although testing this numerically is difficult for larger values of $N$, 
%we propose that, for any system, the Gibbs and Boltzmann entropies will differ for large $N$ (as 
%seen in Fig.~\ref{fig:one}).  
%
%For larger energies at fixed $N$, $\epsilon>N-1$, the right side of
%Eq.~(\ref{eq:omegalowepsilon}) overestimates the number of
%microstates. 
%
%  This implies, as seen in
%Fig.~\ref{fig:universalfigure}, that the Gibbs and Boltzmann entropies
%deviate from the universal curve  for $\epsilon>N-1$ as can be seen in
%Fig.~\ref{fig:universalfigure}.  
%

\section{Concluding Remarks}
We have investigated the Gibbs ($S_G$) and Boltzmann ($S_B$) entropies for a system of $N$ fermions in a one-dimensional
harmonic oscillator potential with total energy $E$, finding that the Gibbs and Boltzmann entropy and
temperature definitions approximately agree at small $E$ while diverging from each other at large $E$. 
In the large energy regime, we find that the corresponding Boltzmann temperature is much higher than
the Gibbs temperature, with the latter being close to $T_{GC}$, the temperature expected based on the
grand-canonical ensemble.  Thus, we find a striking (and potentially experimentally observable) difference
between the Gibbs and Boltzmann pictures for the entropy and temperature in the microcanonical ensemble.  

For sufficiently large $N$, standard thermodynamics arguments imply that the difference between $S_G$
and $S_B$ should vanish.  We found that the 
agreement between $S_G$ and $S_B$ at low energies is connected to the existence of  universal formulas for these entropies that apply 
for sufficiently small $\epsilon = E - E_{\rm min}$, allowing us to numerically establish agreement among these
entropy and temperature definitions for larger $N$.

We proposed that, for any fixed $N$, this agreement will
break down for larger system energies (beyond the universal regime).  This would imply that any system 
at fixed $N$ would exhibit the qualitative behavior shown in Figs.~\ref{fig:one} and \ref{fig:two} for
sufficiently large $E$, with the difference between $S_B$ and $S_G$ (and between $T_B$ and $T_G$) increasing
with increasing $E$.  Since establishing this is quite numerically intensive (except for the small $N$ cases
presented here), we leave further investigation of this issue for future work.

\smallskip
\noindent

{\it Acknowledgments\/} 
%We gratefully acknowledge useful discussions with ... . 
 This work was supported through the REU Site in Physics \& Astronomy 
(NSF grant 1560212)  at Louisiana State University and by NSF grant DMR-1151717.
%--------------------------

\end{document}